# SPECTROPOLARIMETRY OF PRIMITIVE PHOTOTROPHS AS GLOBAL SURFACE BIOSIGNATURES.

William B. Sparks[1,2], M. Niki Parenteau[3,4], Robert E. Blankenship[3,5], Thomas A. Germer[6], C.H. Lucas Patty[7,8], Kimberly M. Bott[3,9], Charles M. Telesco[10], Victoria S. Meadows[3,11]

[1]SETI Institute, Mountain View, CA 94043; [2]Space Telescope Science Institute, 3700 San Martin Drive, Baltimore, MD 21218, USA; [3]Virtual Planetary Laboratory, University of Washington, Seattle, WA 98195; [4]NASA Ames Research Center, Moffett Field, CA 94035; [5]Departments of Biology and Chemistry, Washington University in St. Louis, St. Louis, MO 63130; [6]National Institute of Standards and Technology, Gaithersburg, MD 20899; [7]Institute of Plant Biology, Hungarian Academy of Sciences, Szeged H-6726, Hungary; [8] Space Research and Planetary Sciences, University of Bern, 3012 Bern, Switzerland; [9]Dept. of Earth and Planetary Sciences, University of California, Riverside, CA 92521; [10]Department of Astronomy, University of Florida Gainesville, FL 32611; [11]Department of Astronomy, University of Washington, Seattle, WA 98195.

## ABSTRACT

Photosynthesis is an ancient metabolic process that began on the early Earth, offering plentiful energy to organisms that utilize it, to the extent that they can achieve global significance. The potential exists for similar processes to operate on habitable exoplanets and result in observable biosignatures. Prior to the advent of oxygenic photosynthesis, the most primitive phototrophs, anoxygenic phototrophs, dominated surface environments on the planet. Here, we characterize surface polarization biosignatures associated with a diverse sample of anoxygenic phototrophs and cyanobacteria, examining both pure cultures and microbial communities from the natural environment. Polarimetry is a tool that can be used to measure the chiral signature of biomolecules. Chirality is considered a universal, agnostic biosignature that is independent of a planet's biochemistry, receiving considerable interest as a target biosignature for life detection missions. In contrast to preliminary indications from earlier work, we show that there is a diversity of distinctive circular polarization signatures, including the magnitude of the polarization, associated with the variety of chiral photosynthetic pigments and pigment complexes of anoxygenic and oxygenic phototrophs. We also show that the apparent death and release of pigments from one of the phototrophs is accompanied by an elevation of the reflectance polarization signal by an order of magnitude, which may be significant for remotely detectable environmental signatures. This work and others suggest circular polarization signals up to ~1 % may occur, significantly stronger than previously anticipated circular polarization levels. We conclude that global surface polarization biosignatures may arise from anoxygenic and oxygenic phototrophs, which have dominated nearly 80 % of the history of our rocky, inhabited planet.

1. ## INTRODUCTION

The discovery of thousands of exoplanets, and the consequent implication that there are large numbers of habitable environments throughout the Galaxy, raises the fundamental question, "Are these planets in fact inhabited?" Efforts to resolve this culturally profound and scientifically vital issue are driving an explosive growth in exoplanet research and technology development, aimed at directly imaging and characterizing Earth-like exoplanets. Ancillary research into the origins of planetary systems and comparative planetology to establish an accurate context for understanding our place in the habitable and uninhabitable Universe is likely to represent a major endeavor in the coming decades.

In order to determine whether or not an exoplanet hosts life, we expect to rely on remotely detectable biosignatures – observational signals that are produced by living organisms that can be seen from astronomical distances with suitable telescopes. Ideally, these signatures would be uniquely produced by a biological agent, but in practice, we need to account for the probability of abiotic production (Des Marais et al. 2008; Schwieterman et al. 2018). While much emphasis has been placed on spectroscopic characterization of planetary atmospheres, derivation of atmospheric composition and its interpretation in terms of a biosphere requires observation over a wide wavelength range, extensive modeling, and elimination of numerous potential "false-positive" signals (e.g., Meadows et al. 2018).

A more direct approach is afforded by the concept of "surface biosignatures," where an observable characteristic of the directly-observed biological material serves as a biosignature, with the likelihood of reduced model dependence (Schwieterman et al. 2018). Photosynthetic utilization of the host star's radiant energy is by necessity a surface phenomenon. Furthermore, the interaction of biological material with the incident light through the intermediary of its pigments is a strong interaction rendered polarization-sensitive by chiral structures within the pigments and pigment complexes (Sparks et al. 2009a, b; Patty et al. 2018; Schwieterman et al. 2018). Hence, photosynthetic communities, which have the potential to achieve global environmental dominance, also have the potential to exhibit surface biosignatures through their spectral and circular polarization characteristics. That is, circular polarization spectroscopy is potentially a powerful surface biosignature.

Furthermore, the uniquely biological phenomenon of homochirality is considered a universal biosignature that is independent of a planet's biochemistry (Wald, 1957; Sparks et al. 2009a; MacDermott 2012; Patty et al. 2018; Glavin et al. 2019). A molecule is considered chiral if it is not superimposable on its mirror image. Abiotic chiral molecules exist, but the abiotic synthesis of equal proportions of each "handedness" of the molecules leads to what's termed a racemic mixture. In contrast, biological molecules and polymers are built from a single handedness or enantiomer (e.g., left-handed amino acids and right-handed sugars). Members of each of the biological macromolecular groups – proteins, carbohydrates, nucleic acids and lipids – are chiral and in practice, complex biological molecules typically contain multiple chiral centers. Here, we use the term "chirality" to cover the complexity of homochiral phenomena exhibited in biological systems, from the individual molecular level to supramolecular assemblies. It is an open question in origins of life research how enantiomeric excess and homochirality first arose, but it is considered a universal or agnostic trait of all biochemical life.

The individual molecules, as well as polymers and macromolecules composed of unique enantiomers of their constituent monomers, are typically optically active and rotate polarized light in either a clockwise or counterclockwise direction, dependent on wavelength (optical rotatory dispersion, ORD), or preferentially absorb left or right circularly polarized light (circular dichroism, CD). Polarization signatures offer a well-known probe of biomolecular structure (Kelly

& Price, 2000), and have been shown to be present in pure microbial cultures (Sparks et al. 2009a, b), and in more complex structures such as vegetation and algae (Patty et al. 2017; Patty et al. 2019). With the capacity for circular polarization spectroscopy to detect chirality, without requiring molecular specificity of many biologically motivated life-detection tests, circular polarization spectroscopy therefore has the potential to be a generic, agnostic (universal) biosignature.

It appears likely that circular polarization biosignatures are free from abiotic mimics (Sparks et al. 2009a, b), and can be distinguished from circular polarization originating from multiple scattering effects by clouds or other aerial particles, which create a weak signal that is spectrally very flat (Rossi & Stam 2018). Distinctive polarization features due to biological pigments and pigment complexes potentially allow us to probe exoplanet environments across, essentially, the entirety of their living history with a plausibly pure biosignature, free from many of the model dependencies of atmospheric and surface spectroscopy.

The circular polarization spectrum can provide a remote sensing and *in situ* capability to infer the presence of chiral biological material. For life detection missions, finding an enantiomeric excess $\geq$20 % of one handedness suggests the presence of a biological source (Neveu et al. 2018; Glavin et al. 2019). Acquisition of circular polarization spectra is likely to include incidentally both precision spectroscopy and linear polarimetry. High signal-to-noise telescopic spectroscopy can reveal classical atmospheric biogenic gases, while linear polarimetry will characterize clouds, rainbows, liquids, solids and aerosols in the scattered light of the planet. Taken all together, such observations would provide an effective diagnostic suite to answer the difficult question of whether a particular remote target hosts life, once large, capable, probably space-based, telescopes become available. Within the Solar System, both remote sensing and *in situ* measurements are feasible and targets such as the plumes and surfaces of icy moons, impact craters on Mars, the organic haze of Titan's atmosphere and the clouds of Venus offer example destinations.

Indeed, chirality detectors have already flown within the Solar System, albeit victim of unrelated technical malfunctions (McKay et al. 1998; Goesmann et al. 2015), while new ones are being prepared for upcoming missions (Goesmann et al. 2017). The search for homochirality has also been explicitly proposed for extrasolar life detection (MacDermott 1998).

In this work, we characterize the spectropolarimetric surface biosignatures associated with a variety of anoxygenic phototrophs and oxygenic phototrophs (cyanobacteria). Anoxygenic photosynthetic bacteria represent evolutionary precursors to the oxygenic photosynthetic cyanobacteria, deriving and storing energy from sunlight, synthesizing ATP by cyclic photophosphorylation without the production of oxygen. Anoxygenic photosynthetic microbes have been present for approximately 80 % of the history of life on Earth and hence provide an attractive target to probe epochs prior to the rise of biogenic atmospheric gases, as well as a surface biosignature subsequently. Blankenship (2014) reviews the evolution and history of photosynthesis. The major pigments of photosynthesis, bacteriochlorophylls (Bchls) and chlorophylls (Chls), are optically active molecules with chiral centers, illustrated for example in Senge et al. (2014), their Fig. 6. These pigments absorb in the visible and near-infrared (NIR) from approximately 370 nm to 1040 nm, and are well suited to absorbing light from a variety of stellar types (e.g., G stars and M stars). In addition to the intrinsic chirality of the pigments, larger conformational pigment and pigment-protein complexes can be chiral. An interesting phenomenon arises from large and dense chiral supramolecular assemblies, such as can be found in vegetation and algae. Such chiral aggregates can cause very intense polarization signals with non-conservative anomalously shaped bands that can extend beyond the molecular absorption bands (Keller & Bustamante 1986; Garab & van Amerongen 2009). The characterization of samples of such organisms offers encouragement that surface biosignature exploration of distant planets may be possible for a large fraction of their history and that for both extrasolar planets and within the Solar System, free from terrestrial assumptions and specificity, we may make significant progress in agnostic life detection using circular polarization spectroscopy.

Section 2 presents our methodology, section 3 the empirical results, section 4 discusses the measurements and their relevance to life detection, and section5 concludes.

## 2. METHODS

### 2.1 Samples

We measured the *in vivo* absorbance spectra and linear and circular polarization spectra of pure cultures of photosynthetic microbes and environmental photosynthetic microbial mats. Pure cultures of representatives of the five major anoxygenic photosynthetic groups were analyzed in this study (Table 1, Figure 1). The recently described sixth photosynthetic group including *Chloracidobacterium thermophilum* was not analyzed (Bryant et al. 2007). Cultures were obtained from the Blankenship lab, which maintains them for various pigment and reaction center studies. Many of the anoxygenic phototrophs are highly characterized type strains. *Chloroflexus aurantiacus* J-10-fl was obtained from Beverly Pierson (Pierson & Castenholz, 1974), and is maintained at NASA Ames Research Center by Parenteau.

On the basis of its differing absorption and reflection polarization spectra, §3.1.3 below, *Chlorobaculum tepidum* was selected for further study and an increased sample size of the pure culture was harvested at the exponential phase from the Blankenship laboratory and shipped to NIST following standard protocols in January 2016 (the "Jan16" sample). When we attempted to measure the Jan16 sample, we experienced a hardware problem with the polarimeter, requiring a period of several months for recovery of the instrument. Following replacement of the internal monochromator, the instrument was brought back into service and was found to function to the same level of accuracy as prior to the failure. Measurements of the Jan16 sample eventually commenced during May 2016. For the intervening four months, the samples were stored in a freezer at NIST. When thawed out these samples exhibited clear indication of cell lysis leading to the extracellular release of pigments.

Samples of photosynthetic microbial mats were obtained from Yellowstone National Park under Parenteau's permit #1549. These laminated mats contain cyanobacteria in the surface layer, followed by underlayers of anoxygenic phototrophs and chemotrophs (Table 2, Figure 2). The mats contain genera represented by the pure cultures (e.g., *Chloroflexus* and *Roseiflexus*), and the pure cultures were used to help deconvolve the polarization signals of the complex natural mats. The mats were harvested in the field, immediately placed on dry ice, then stored at -80 ºC in the laboratory, and then shipped on dry ice to NIST for the circular polarization measurements.

Following freezing, storage and transportation to NIST – they were not measured *in situ* – the mats were measured using the same configuration as that of the reflection measurements for the pure cultures.

### *2.2 Pigment analyses*

The *in vivo* absorption spectra of the pure cultures and photosynthetic microbial mats were obtained by sonicating the cells in cold Tris-sodium-magnesium (TSM) buffer. The cohesive mats were first dispersed using a glass tissue homogenizer prior to sonication to ensure total disruption of the cells to release the pigment-protein complexes. After sonication, the cellular debris was removed by centrifugation. The absorption of the pigment-protein complexes was measured from 350 nm to 1100 nm on a Shimadzu UV-1601 scanning spectrophotometer (Pierson & Parenteau, 2000). The purpose in performing the *in vivo* pigment analyses independent of the polarization measurements was to identify the major absorption peaks using traditional microbiological methods to help deconvolve the spectral features in the circular polarization spectra.

### *2.3 Measurement of the polarization spectra*

#### *2.3.1 Polarimetric instrumentation*

The polarimeter and measurement methods are described in detail in Sparks et al. (2009a). The instrument is a Hinds Instruments, Series II/FS42-47 dual photoelastic modulator (PEM) full Stokes polarimeter optimized for measurement of circular polarization in the presence of significant linear polarization. The polarimeter has spectral resolution of approximately 10 nm full width half maximum (FWHM) and covers the spectral range 400 nm to 800 nm by scanning in 5 nm steps. The polarization is described by the Stokes vector $S \equiv (I, Q, U, V) \equiv I(1, q, u, v)$ where $I$ is the total intensity, $Q$ and $U$ are the unnormalized linear Stokes parameters, $V$ is the unnormalized circular polarization Stokes parameter, and $q$, $u$, and $v$ are the normalized linear and circular Stokes parameters. The degree of linear polarization is $p_l = \sqrt{q^2 + u^2}$, the linear polarization position angle is $\zeta = \frac{1}{2} tan^{-1}(u/q)$ and the degree of circular polarization is $v$. (The degree of polarization is 1 or 100% when perfectly polarized.) The target performance of measuring a degree of circular polarization $v = 10^{-4}$ in the presence of linear polarization degree $p_l = 0.03$ is achieved. Since the linear and circular polarization data are carried on the 2$f$ and 1$f$ modulation frequencies of the PEM resonant frequency $f$, respectively, there is no cross-talk between the linear and circular polarization measurements.

#### *2.3.2 Sample measurement methods*

Each sample of the pure cultures and photosynthetic microbial mats was placed in a shallow Pyrex glass Petri dish, with examples illustrated in Figure 1. The polarization spectra of the pure cultures were acquired in both a reflection and transmission mode, and were normalized to a reference white light spectrum obtained by using the same illumination source as the samples, an adjustable 150W stabilized quartz-halogen fiber optic illuminator (Dolan-Jenner DC950). The polarization spectra of the photosynthetic microbial mats were acquired in reflection mode. The polarization was corrected for the (small) baseline polarization of the white light. No significant polarization was introduced by the Petri dish. These two configurations are described in more detail in Sparks et al. (2009a; b). In the reflection mode, unpolarized light exiting an integrating sphere illuminates the sample from above and is reflected or scattered directly into the polarimeter, which views the sample from above with an unobscured optical path. In the transmission mode, the illumination source is directed to a white plaque beneath the sample, and the light from this white scattering surface then passes through the sample and, as in the reflection measurement, from the sample directly into the polarimeter. Except where explicitly stated, the reflection measurements used a black mat background beneath the sample so that the reflected light was dominated by scattering solely from the microbial sample. We measured a number of the samples at a range of optical depths, and determined that the signal in both transmission and reflection was robust against such variations. Once an optical depth of order unity is obtained, the polarization signal strength is roughly constant (Sparks et al. 2009b).

## 3. RESULTS

### *3.1 Pigment and Polarization Spectra: Pure Cultures*

Schwieterman et al. (2018) present the absorbance spectra and characteristic wavelengths of features in the spectra of pigments (their Fig. 9, Table 1). Figure 3 presents similar spectroscopic and polarimetric data for the filamentous anoxygenic phototroph (FAP) *Chloroflexus aurantiacus* showing the circular polarization transmission and reflection spectra, the absorbance of the pigments in the intact cells measured by the polarimeter, and the *in vivo* spectrum of the pigment-protein complexes isolated from the cells, which was measured independently. The polarimeter absorbance spectrum is included to enable identification of pigment absorption maxima and correlate them to the pigment polarization features.  The polarimeter generates clear absorption spectra while in transmission mode, but yields less structured spectra while in reflection mode. Therefore, we measured the *in vivo* absorption spectra using a different method and instrument to corroborate the polarimeter absorption measurements, and in particular to help correlate and interpret pigment polarization features in reflection mode.  This illustrates that (i) the absorption features of the whole cells measured by the polarimeter closely agree with the *in vivo* spectrum, and (ii) the circular polarization features correlate with these absorption peaks. These are generic characteristics common to all samples, and the polarimeter absorption spectra are subsequently plotted in Figures 4 and 5. Figure 3 identifies specific pigment absorption maxima and correlates them to polarization features, which recur in some or all of the other samples (see also Schwieterman et al. 2018).

Figures 4 and 5 present the results of the measurements for the pure cultures of the selected anoxygenic phototrophs, derived as described in the Methods section, with the upper row giving the transmission circular polarization spectra, the middle row the reflection circular polarization spectra and the bottom row the absorbance spectra as measured by the polarimeter. The uncertainties shown with the data represent the standard deviations of the mean of multiple repeated measurements. Below, we discuss each of the samples in turn, roughly in decreasing order of the strength of the polarization signal.

*Chloroflexus aurantiacus:*
Figures 3(a) and 3(b) and 4(a) and 4(b) show the circular polarization spectra for transmission and reflection, respectively, for the pure culture of *Chloroflexus aurantiacus*, a filamentous anoxygenic phototroph (FAP). The transmission spectrum shows a large, broad polarization peak of approximately 0.2 % polarization at about 530 nm. The shoulder of this prominent peak in the polarization spectrum may correspond to the $Q_x$ absorption band of bacteriochlorophyll *c* (Bchl *c*), visible as a small peak in the absorption spectrum. The carotenoid absorption in the blue does not correspond to any strong polarization feature. At wavelengths longer than 700 nm, there is a sharp rise in the circular polarization spectrum to a peak of 0.5 %, at about 760 nm, with a dip on the rising slope at about 740 nm, corresponding to the strong $Q_y$ absorption band of Bchl *c*. In the reflection polarization spectrum, the $Q_x$ features are absent. However, there is a very large, clean negative circular polarization feature, extending from 700 nm to 800 nm, with peak (absolute) value also at about 740 nm, which is, within the uncertainties, at the location of the Bchl *c* absorption maximum as shown in Figure 4(c). In the blue part of the spectrum, there are indications of additional weak polarization structure.

*Rhodobacter capsulatus:*
Figures 4(d) and (e) show the transmission and reflection polarization spectra of *Rhodobacter capsulatus*, a purple non-sulfur anoxygenic phototroph. The main pigment utilized by this organism is Bchl *a*, which has absorption peaks at about 805 nm and 860 nm, and there is a monosignate negative polarization feature in the transmission spectrum at about 800 nm. There is also a relatively strong polarization peak at about 510 nm, potentially associated with carotenoids with polarization levels approximately 0.1 %. By contrast, these features are absent from the reflection spectrum, where the significant (albeit weak) polarization peak is at an inconspicuous region of the transmission spectrum from 600 nm to 700 nm. The *in vivo* pigment absorbance spectrum (not shown) has a small maximum at 675 nm, but this peak is absent in the polarimeter-based absorption spectra shown in Figure 4 (f).

*Chlorobaculum tepidum*
Figures 4(g) and (h) show the transmission and reflection polarization spectra of *Chlorobaculum tepidum*, a green sulfur anoxygenic phototroph. In transmission, there is an extremely prominent antisymmetric polarization feature from 700 nm to 800 nm, with a sign reversal at about 750 nm, coinciding with the maximum $Q_y$ absorption of Bchl *c*, visible in Figure 4(i). The maximum degree of polarization for this highly distinctive profile is about 0.25 %. In the blue region of the spectrum, carotenoids or the Soret band of Bchl *c* cause additional polarization structure. By contrast, in reflection, the overall polarization level is much reduced. The characteristic S-shaped curve of the transmission spectrum is absent, with instead, broad, weak polarization peaks to the red and to the blue. We also analyzed *C. tepidum* after subjecting the culture to freeze-thaw and lysis of the cells and release of chlorosomes and pigments, which shows that the polarization increases by an order of magnitude (see below for a full discussion).

*Roseiflexus castenholzii:*
Figures 5(a) and (b) show the transmission and reflection polarization spectra for *Roseiflexus castenholzii,* another FAP. In contrast to *Chloroflexus*, the overall level of polarization is very weak, reaching at most 0.04 %. The *in vivo* absorption spectrum (not shown) reveals absorption of bacteriochlorophyll *a* (Bchl *a*) at 810 nm and 885 nm, which is mostly beyond the range of this polarimeter. However, the absorption maxima at about 810 nm, shown in Figure 5(c) appears to be introducing rapid changes to the polarization spectrum in both transmission and reflection. Elsewhere there are small peaks and dips in the polarization degree, potentially correlating with the $Q_x$ band of Bchl *a* and/or carotenoids.

*Thermochromatium tepidum:*
Figures 5(d) and (e) show the polarization spectra for *Thermochromatium tepidum*, a purple sulfur anoxygenic phototroph, and the absorbance spectrum is shown in Figure 5(f). The transmission polarization spectrum is unusual in being entirely negative, with a flat, broad (negative) peak in polarization degree blueward of about 480 nm. This may be another example of a broad S-shaped feature, from about 550 nm to the blue, corresponding to the prominent broad carotenoid absorption band. Additional polarization structure at about 800 nm is likely caused by the onset of the Bchl *a* bands, which continue into the near NIR. In reflection, the primary feature is also towards the NIR part of the spectrum, although its amplitude is low.

*Heliobacterium modesticaldum:*
Figures 5(g) and (h) show the transmission and reflection polarization spectra for *Heliobacterium modesticaldum,* a heliobacterium. In the transmission spectrum, a sharp peak around 780 nm is followed by a sign reversal and decrease continuing to the red edge of the detector range of sensitivity. The principal pigment associated with this organism is Bchl *g*, which has $Q_y$ absorbance maxima at about 674 nm and 788 nm, shown in Figure 5(i), where polarization structure is evident, particularly at the red extreme of the spectrum. There is additional weak polarization structure across the entire spectrum in transmission. In reflection, only very weak polarization signatures are seen, with a weak feature at 570 nm which corresponds to the $Q_x$ absorbance band as shown in Figure 5(i). A sharp rise at the blue edge of the spectrum at about 400 nm appears to be significant.

*Prosthecochloris aestuarii*
Figures 5(j) and 5(k) show the polarization spectra for *Prosthecochloris aestuarii,* also a green sulfur anoxygenic phototroph. In transmission, a sharp negative peak can be observed at 445 nm and a clear S-shaped feature is present around 685 nm. The polarization level for the reflected samples is very weak, and largely featureless. Interestingly, even though the microbe utilizes the pigments Bchl *c, a* and Chl *a* (Table 1), there is only a strong Chl *a* absorption band present at about 680 nm (see Figure 5(l)); with a corresponding polarization feature (Figure 5(j)).

### 3.2 Pigment and Polarization Spectra: Photosynthetic Microbial Mats

Table 2 lists the major phototrophs and pigments present in the six photosynthetic microbial mats. Each mat typically contains two to three major photosynthetic groups, but these are complex communities containing thousands of species of microbes. Parenteau et al. (in preparation) will present reflection spectra of field samples in depth. Figure 6 shows a first attempt to ascertain the circular polarization levels from light scattered by such photosynthetic mats. In general, most mats did not yield a strong reflectance signal; however the Chocolate Pots *Synechococcus*-Chloroflexi mat, Figures 6e and 7c, shows a significant and distinctive polarization maximum at 740 nm, which corresponds to the absorption maximum of Bchl *c* in *Chloroflexus*. While this feature is clearly associated with Bchl *c*, the sign is reversed relative to the pure culture. Such a sign reversal can occur at a specular reflection, though it is unclear where such a reflection might be occurring, or if we may be seeing a polarization-sensitive surface scattering/reflection phenomenon. The relative contributions of specularly reflected light and internally scattered light (where polarization can self-cancel) must differ between the pure culture and the more natural environment of the mat. There are other polarization structures within the spectra, at levels within the range of our precision polarimeter, and almost certainly associated with the pigmentation of the microbes in the mats, however in an absolute sense these polarization levels are low.

In addition, signals from membrane assemblies, such as the different architectures of intracytoplasmic membranes in purple bacteria, and chlorosomes in green sulfur and filamentous anoxygenic phototrophs, can yield mixtures of signal shapes and signs resulting simultaneously from different levels of structural organization.

We have presented polarization spectra of pure cultures and of microbial mat communities that include some of the species discussed individually above. To analyze the three perspectives in combination, Figure 7 displays a pure culture's (*Chloroflexus aurantiacus*) transmission and reflection, which may be considered as end member spectra contributing to the reflection of a microbial mat that contains a relatively high fraction of these phototrophs, and an example of such a mat. We observe a feature at 740 nm in the pure culture's reflected polarization spectrum. The same feature occurs with the sign reversed in both the pure culture's transmission and in the microbial mat reflection. Hence, we are able to trace a spectral polarization feature in an environmental sample to a specific biological source through the pure culture's end member spectra. This ability to validate the biological nature of circular polarization signature – and techniques to deconvolve environmental spectral – will be significant for life detection where the biosignature may arise from complex communities.

### *3.3 Enhancement of circular polarization signal upon cell lysis*

In our earlier work, we found that the reflection and transmission spectra of cyanobacteria were very closely related (Sparks et al. 2009a, b). In contrast, the measurements in this study reveal significant differences in the two types of spectra. We therefore selected an extreme example of the discrepancy, as given by *Chlorobaculum tepidum*, contrasting Figure 4g and Figure 4h, for further study. The striking S-shaped or derivative curve of the transmission spectrum, Figure 4g, was not evident in the reflection experiment, Figure 4h, in stark contrast to the cyanobacteria results of Sparks et al. (2009a). Hence, we obtained additional samples of *Chlorobaculum tepidum* cultures for this study (the "Jan16" sample) but were unable to measure their polarization spectra for approximately four months due to a hardware failure. During this period the samples were stored in a freezer at NIST and when thawed out, demonstrated clear indications of cell lysis and release of pigments into the media, see §2.1.

Figures 8 (a) and (b) show the polarization spectra obtained in March 2015 for the original fresh culture, and the measurements obtained in May 2016 for the Jan16 culture following a four month storage period at -20 ºC, Figures 8 (c) and (d). Figures 8 (e) and (f) show polarization spectra of the Jan16 sample obtained a full two years after the initial measurement (storage -20 ºC), obtained February 2018.

Figures 8 (a) and (b) replot the data of Figures 4(g) and 4(h), now on the same scale. The transmission polarization spectrum of *C. tepidum* is extremely distinctive (especially the reversed-sign feature straddling 750 nm corresponding to Bchl *c*). Despite this, the original reflection spectrum of the presumably healthy pure culture showed almost no polarization apart from a weak, low level rise seen expanded in Figure 4(h). In 2016 the transmission spectrum showed a substantial change in the strength of the positive side of the double polarization feature: the hitherto positive side of the reversed-sign polarization feature (to the blue of 750 nm) became negative, while the negative side (to the red) did not show much change initially, before weakening. This is reminiscent of the type of changes observed in decaying leaf matter (Patty et al. 2017) where the negative and positive chlorophyll bands evolve independently. Interestingly, at this time the reflection spectrum from the lysed *C. tepidum* sample showed a very strong polarization feature from 400 nm to 500 nm (corresponding to the Soret absorption band of Bchl c), about an order of magnitude amplification relative to the previously existing weak feature of the healthy culture. In 2018, after two years of storage at -20 ºC and further cell lysis, the transmission spectrum continued its evolution; the polarization feature from 750 nm to 800 nm decreased in amplitude and reversed sign. Meanwhile, the reflection spectrum developed a second extremely prominent polarization feature from about 720 nm to 800 nm at the same wavelengths as the strong polarization feature in the transmission experiment, which is associated with the $Q_y$ transition of Bchl *c*. Again, this is an increase of about an order of magnitude compared to the intact, healthy sample. We discuss reasons why this remarkable enhancement of polarization may have occurred in the Discussion section.

## 4. DISCUSSION

### 4.1 Spectropolarimetry of pure cultures and photosynthetic mats

We have undertaken a systematic study characterizing the circular polarization spectra of pure cultures of anoxygenic phototrophs, and we used these spectra to help deconvolve the spectra of complex photosynthetic microbial mat communities in the natural environment. We began with measuring the *in vivo* absorption spectra of isolated pigments and pigment-protein complexes from each culture, and used that data to help interpret the circular polarization transmission and reflectance spectra of the cultures, and then the microbial mats. Knowing the characteristics of the constituent pigments enables us to identify features in pure microbial cultures, and knowing the features of individual microbial species allows us, in principal, to understand the composite spectra of microbial mat communities. This was demonstrated in Fig. 7 where we were able to trace the Bchl *c* present in the pure culture of *Chloroflexus* into the complex environmental *Synechococcus*-Chloroflexi photosynthetic mat.

Sparks et al. (2009a; b) presented the first extensive study on oxygenic photosynthetic microbes from a similar (astronomical) perspective, in which it was found that the polarization spectra of a pure culture of cyanobacteria in reflection mimicked those in transmission, apart from the squared absorption amplitude resulting from the double pass through the culture. The polarization features correlated well with the spectral features, and both monosignate and derivative-shaped curves were found associated with the chlorophyll *a* pigment and accessory pigments phycocyanin and phycoerythrin. From those initial cyanobacteria pure culture results, a simple heuristic explanation of the relationship between polarization in reflection and transmission was presented, namely that optical depth dominated in both cases. It was supposed that for a reflection experiment, the incoming light beam penetrated to a multiple scattering surface defined by unit optical depth, below which polarization information was lost, but that by necessity it re-emerged through the layer between unit optical depth and zero optical depth in a fashion analogous to the transmission experiment, resulting in similar polarization spectra.

Here, in a second major study of microbes encompassing a more diverse suite of pigments, phototrophs, and environmental microbial mats, we find an extraordinary range of polarization properties and interrelationships between reflection, transmission, and absorption. A conclusion of this study is that there remains a great deal more to understand and learn than the early indications from the cyanobacteria pure culture led us to believe.

Unlike the previously studied pure cultures, here we found little correspondence between the reflected and transmitted spectra. In some cases, the polarization spectra were completely different, and in others, features from one pigment were present in both, while other pigments were not, challenging the task of global prediction of signal strength from, for example, the circular dichroism (CD) literature. There were transmission features absent in the reflection spectra, though conversely, only a small number of features in the reflection spectra did not appear in the transmission spectra – the unidentified reflection feature from 600 nm to 650 nm in *Rhodobacter capsulatus* and the spike at the blue edge of the *Heliobacterium modesticaldum* spectrum. Interestingly, in some cases the features are present in both reflection and transmission spectra, but with large differences in signal magnitude between them. Some of these differences are well-illustrated by the spectra of *Chloroflexus aurantiacus*, Fig. 7, which seems to have many more features identifiable in transmission than in reflection. In reflection, a single strong band dominates, with shoulders at 730 nm and 765 nm, apparently correlating with features present in transmission.

Nevertheless, we are able to identify individual biological elements within complex natural microbial communities, and to potentially deconvolve environmental spectra, including polarization, into their constituents.

The cause of these differences remains unresolved, with candidate effects ranging from the internal cell structure and arrangement of photosynthetic apparatus, to macroscopic scattering phenomena external to the microbes. A polarization sign reversal on reflection could, in principle, null the dichroic effect of the medium on the beam as it propagates back through the sample. Other factors could relate to the geometry of the pigments relative to the cells and their membranes or internal cell structure (e.g., pigments arranged in chlorosomes and chiral domain organization, Patty et al. 2018), which we discuss below.

**4.2 Chirality and structural organization of pigment systems**

Historically, CD measurement of the photosynthetic apparatus of microbes has taken place on isolated fractions, and is dependent on the surrounding chemical environment (including use of solvent), concentration, and temperature. Our circular polarization measurements were performed on whole cells and complex communities of microbes, which is more relevant for treatment of chirality as a surface biosignature for life detection. These polarization spectra represent a mixture of the features, signs, and magnitude associated with pigments and pigment complexes at several layers of structural organization in whole cells, as well as in multi-layered photosynthetic mats. The pigments and the structural arrangements of the pigments vary considerably among the different photosynthetic microorganisms, and there appears to be a corresponding remarkable variety of circular polarization spectra. In addition, as noted above, polarization spectra are additive, and a variety of signals interact to yield net spectra for each experimental measurement. As such, this renders the interpretation of measurements on environmental samples more complicated. We have endeavored to identify wherever possible the biophysical reasons for the resultant spectra (noting that from a life detection perspective, this is not expected to be critical).

In general, there are three levels of structural organization that can generate CD signals: (1) photosynthetic pigments are intrinsically chiral as they are composed of asymmetric carbon atoms, (2) excitonic coupling between two or more pigment molecules (where the excited state is delocalized over a few pigments), and (3) larger macromolecular pigment-protein complexes and membrane assemblies (Garab & van Amerongen, 2009, and references therein). CD studies of monomeric solutions of chlorophyll reveal that the substituents at the asymmetric carbon numbers 10, 8, and 7 control the sign and magnitude of the spectra (Houssier & Sauer, 1970), which are generally thought to reflect the shape of the pigment absorption maxima. However, due to the planar geometry of photosynthetic pigments, these intrinsic signals are generally weak, and cannot be identified in our transmission and reflection spectra. The signals arising from excitonic coupling are stronger, and have a characteristic sinusoidal shape with positive and negative bands. The Bchl *c* in *Chlorobaculum tepidum* provides an example of this in Fig. 4c. The signals from larger chiral polymer and salt-induced (psi-type) aggregates yield non-conservative anomalously shaped bands. These can extend beyond the molecular absorption bands (Keller & Bustamante 1986; Garab & van Amerongen 2009). In addition, signals from membrane assemblies, such as the different architectures of intracytoplasmic membranes in purple bacteria, and chlorosomes in green sulfur and filamentous anoxygenic phototrophs, can yield mixtures of signal shapes and signs resulting simultaneously from different levels of organization.

The effect of the structural organization of the pigments on the shape and magnitude of the circular polarization is well illustrated by the course time measurements on *Chlorobaculum tepidum* (Figure 8). Circular polarization measurements on chlorosomes generally show a large diversity, resulting from two overlapping CD spectra (Griebenow 1991). Within the chlorosomes, the pigments are self-coordinated rather than being oriented by the pigment-protein environment. Likely, upon lysis of the cells, the pigments have auto-assembled into aggregates larger than those in the chlorosomes, and the spectral shape in transmission corresponds with reported size-dependent changes in CD (Prokhorenko 2003). It is more difficult to directly explain the much larger signal in reflection, similar to the observed large differences in the chlorosomes of *Chloroflexus aurantiacus*. Likely other factors play a large role in the differences between transmission and reflection polarization measurements on chlorosomes.

It is highly likely that the transmission and reflection spectra contain mixtures of all signal types, and thus it is often hard to assign specific features to any particular structural organization. Nevertheless, what has emerged from this study is a catalog of biological circular polarization spectra that have no known abiotic mimics, whose magnitude (degree of circular polarization) is stronger that previously appreciated, which helps bolster the potential of circular polarization spectra as a biosignature.

**4.3 Relevance to life detection on Solar System bodies and exoplanets**

The majority of the photosynthetic microbial samples studied here exhibited degrees of polarization in transmission at levels up to $v \cong 0.5\%$ ($5 \times 10^{-3}$). In reflection, the range is about $10^{-4}$ to $10^{-3}$, consistent with previous findings for cyanobacteria and other organisms studied: a modest signal, but still some two orders of magnitude higher than ambient abiotic polarimetric

"noise." In §3.3, we discussed an empirically observed amplification of the signal that exceeds this range. Other studies have found additional examples of higher circular polarization levels (Patty et al. 2019), up to 2 %. It is essential to accumulate a broader understanding of the range of microbial polarization levels present, and empirical data such as Earth observations from space that include both microbial and vegetation dominated scenery would be invaluable for exoplanet surface biosignature modeling and experiment design based on remote detection of chiral signatures.

If a visible light or UV spectropolarimeter is included on a large telescope it may be able to measure the linear and circular polarization of photosynthetic microbes, and characterize chiral biosignatures, even if the pigments differ substantially from those found on Earth. As noted in the introduction, photosynthetic microbes inhabiting continental areas, marine intertidal, and shallow marine settings on exoplanets can provide surface biosignatures for the remote detection of life (Schwieterman et al. 2018). The reflectance spectra of the photosynthetic pigments (e.g., the red edge of chlorophyll *a*, or the near-infrared-edge of purple bacteria [Sanromá et al. 2014]) could be detectable through different cloud coverage levels and atmospheric compositions inferred via direct imaging by next generation space telescopes, such as the Large UV/Optical/IR Surveyor (LUVOIR), which is one of four Decadal Survey on Astronomy and Astrophysics (Astro2020) Mission Concept Studies (LUVOIR Final Report, 2019).

While the reflectance spectra of these features are generally weaker than the transmission spectra by an order of magnitude, they could be observed with roughly 10,000 hours on a LUVOIR-size telescope with a visible light spectropolarimeter, in the case of an inner-edge habitable zone terrestrial super Earth (R ~ 1.6 $R_E$) covered in the pigment. Currently, the proposed LUVOIR-A architecture concept includes a UV wavelength high resolution spectropolarimeter called POLLUX. Estimates for the UV spectropolarimeter POLLUX as described in the LUVOIR Final Report (LUVOIR Final Report, 2019) show that for the 15 m primary architecture A design, a parts-per-million level polarization would be detectable after only a few hours of integration time around bright, nearby (< 50 pc) stars. However, in polarimetry we must consider both precision and the available flux and thus must scale the signal by the anticipated light reflected by the planet, either in combination with that of the star, or isolated from the star but much fainter. Hence it is highly unlikely that the next envisaged generation of telescopes could apply these techniques directly to Earth-like planets around Solar-type stars.

A parts-per-million magnitude signal (like that of the pigments reflectance spectra shown in this study) could, in theory, be detected on a terrestrial super Earth with only ~100 hours of observations with a visible light spectropolarimeter *if* it orbited closely (~0.05 AU) to the star, however such planets are unlikely to harbor life for typical Solar-type stars. (SNR limit estimates are not calculated for a visible light polarimeter for LUVOIR as none are currently planned, so our estimates are based roughly on the instrumental performance of top performing ground-based telescopes, and the precision of POLLUX). It is worth noting that the UV Polarimeter POLLUX can reach SNR within instrumental limits for stars with U band magnitudes brighter than M~18 after a few hours of integration time (see figure 13-5 in the LUVOIR Final Report) and that in general the SNR limits improve towards longer wavelengths.

Both the UV and visible regions of a spectrum however can provide chiral signatures. By utilizing the UV spectral region rather than the visible there is potential access to a plethora of optically active transitions associated with biomolecules such as nucleic acids, proteins and amino acids, including for example, protein secondary structures, such as the α-helix. For remote observing the relative contributions of the star's spectral energy distribution and influence of the presence of an atmosphere should be taken into consideration. While cooler stars in general have lesser amounts of UV flux, this can be compensated for by (a) their habitable zone is closer to the star which improves an exoplanet's contrast ratio relative to the star (b) many faint cool stars are extremely active, and exhibit brief, very luminous UV flares which can illuminate close-in habitable exoplanets. Light echoes of these flares by the exoplanet further improve the contrast ratio due to the time delay for the flare's arrival (Sparks et al. 2018). With a time resolved instrumental capability, such a technique could be available to LUVOIR. Note that the use of light echoes allows access to planets which are very close to their host star, and does not require a coronagraph.

There is an increasing appreciation of aerobiology, although *in situ* microbial communities inhabiting terrestrial aerosol particles and dust are not yet well-established (Smith 2013). The possibility of life in the clouds of Venus has been discussed for decades (Morowitz & Sagan 1967 and many others subsequently). Seager et al. (2021) present a detailed account of a potential Venusian aerial biosphere utilizing a droplet habitat and cyclical, dynamic life cycle. Survival in this environment would prove extremely challenging given, for example, the exceedingly low water activity of the sulphuric acid droplets. Nevertheless, Venus offers a potential Solar System application for an empirical, chiral remote sensing experiment. Conceivably, if an exoplanet's atmosphere were inhabited densely enough by microorganisms, transit spectropolarimetry might be a feasible detection method (Smith 2013) identified by Strassmeier et al. (2014) as a possible Extremely Large Telescope (ELT) key application. In that vein, Judge (2017) proposed spectropolarimetry of the vapor plumes of Europa and Enceladus as they transit Jupiter and Saturn, respectively, as a means to probe for the presence of biomarkers.

Within the Solar System in general, polarimetric methods lend themselves to remote sensing surveys of targets seen in scattered light, such as the vents and deposition blankets of Ocean World plumes, while *in situ* instrumentation could utilize either transmissive or reflective configurations applied to locally harvested samples. Biological pigments can serve to protect against UV radiation (Cockell et al. 2003; Schwieterman et al. 2018), and, coupled to other survival strategies, may be utilized by exotic forms of extant life. Evidence may be sought for preserved chiral signatures in subsurface or otherwise protected environments, while the remains of biological organisms can also retain a chiral signature following recent exposure to a hostile environment, such as ejection in a plume (Hand et al. 2009).

## 5. CONCLUSIONS

We have undertaken a systematic study of increasing complexity, from spectra of purified pigment samples, through pure cultures of intact phototrophs, and from there to complex microbial communities from the natural environment on different mineral/rock substrates. By doing so, we were able to understand the biophysical origin of the signals present in both the spectroscopic and polarimetric data, though differences between reflection and transmission remain unexplained. We found correlations between the polarimetric signatures arising from the chirality of the biomolecules and the corresponding absorption spectra. In addition, *we encountered an unsuspected, enormously rich diversity in almost every characteristic*. We found that, contrary to anecdotal assumptions, the circular polarization signal can be strong; that some but not all pigments exhibit a polarization signature; and that the reflection polarization spectra differ in sometimes major ways from the transmission spectra, unlike those of earlier cyanobacteria studies. The polarization signatures are correlated with spectral absorption features, as expected, and are comprehensible in the context of our knowledge of the implicit pigments and pigment complexes in the organisms, but in ways which differ from organism to organism and pigment to pigment.

Reflectance measurements of the circular polarization spectra of pure cultures and microbial mats from a variety of environmental settings, here and in earlier work, show that biological pigment absorption and circular polarization are strongly coupled. If photosynthesis were to evolve on a habitable terrestrial exoplanet, polarimetric measurements could help interpret spectral data from a future life detection mission, and allow us to infer a biological presence even if pigmented phototrophs evolved to have different absorption maxima given the different stellar spectral irradiance and different atmospheric composition.

We discovered that one of the studied cultures, chosen because its transmission and reflection polarization spectra differed, exhibited an increase in polarization signal strength in reflected light by an order of magnitude as the culture degraded in health and evidence of cell lysis became apparent. For individual measurements, the *reflected* signal reached between 0.5 % and 1 %, far higher than the canonical $10^{-3}$ polarization level characteristic of the reflected light polarization of cyanobacteria (Sparks et al. 2009a). Patty et al. (2019) found in their study of eukaryotic phototrophic organisms, that samples of brown algae *Fucus serratus* and *Fucus spiralis* showed an extremely high degree of circular polarization in transmission, up to 2 % polarization, whereas other algae in their study showed weaker or no signals, which corresponded to a range in polarization properties within the algae of a factor of 1000! We conclude that: (a) the anecdotal "very low signal strength" for polarization in life detection does not always hold and in fact the signals can be very strong, and (b) strong polarization signals can persist after the destruction of the microbial samples.

Hence in the anoxygenic phototroph samples, the signal is present not only for extant, living, healthy microbial communities, but also in decaying cultures with cells breaking apart and, presumably, releasing their pigments into the ambient medium. We do not know how long the dramatic signal enhancement that resulted will persist. In our experiments, it remained high, and was continuing to increase after two years.

The addition of polarimetric information adds a unique and crucial element to the life-detection endeavor, namely, an ability to infer with remote sensing the presence of chirality, one of the most important, pure biosignatures. In an exoplanet context, photosynthesis and in particular anoxygenic photosynthesis have been present for most of the history of terrestrial life, and given its potential to achieve global significance, may reasonably be expected to arise elsewhere and provide a potential biosignature visible for that entire history. The techniques are also applicable for exploration within the Solar System, where there are plentiful photons. The anticipated absence of photosynthesis elsewhere in the Solar System suggests targeting the ultraviolet spectral domain where strong chiral biological signals abound arising from generic biological material such as amino acids and protein secondary structures, such as the α-helix. From an astronomical perspective, the mere presence of a polarization signal is evidence of the presence of chirality and hence life, though from a biological perspective it is important to dig deeper into the underlying biophysics and biochemistry. By doing so, we gain not only knowledge of the hypothetical exospecies, but insight into their biology.

We conclude that we have hit upon a truly remarkable multiplicity of empirical results, defying systematic consolidation, implying that the study of circular polarization as a biosignature is in its infancy. Much more needs to be done to tease out the underlying systematics and causal influences. Empirically, however, the mere presence of circular polarization signatures, sometimes strong, associated with chiral biomolecules bodes well for its use, both *in situ* and in remote sensing, as an important universal, agnostic biosignature.

**ACKNOWLEDGEMENT**

This work was performed as part of NASA's Virtual Planetary Laboratory, supported by the National Aeronautics and Space Administration through the NASA Astrobiology Institute under solicitation NNH12ZDA002C and Cooperative Agreement Number NNA13AA93A, and by the NASA Astrobiology Program under grant 80NSSC18K0829 as part of the Nexus for Exoplanet System Science (NExSS) research coordination network.

## TABLES

**Table 1.** Pure cultures of the five major groups of anoxygenic phototrophs containing bacteriochlorophyll (Bchl) pigments analyzed in this study.

| Pure culture | Photosynthetic group | Pigments | Environmental source |
|---|---|---|---|
| *Thermochromatium tepidum*, Type strain (DSM 3771) | Purple sulfur | Bchl *a* | Continental hot spring |
| *Rhodobacter capsulatus* strain SB 1003 (ATCC BAA-309) | Purple non-sulfur | Bchl *a* | Soil |
| *Prosthecochloris aestuarii* 2K | Green sulfur | Bchl *c, a*; Chl *a* | Marine intertidal lagoon |
| *Chlorobium tepidum* TLS, Type strain (ATCC 49652) | | Bchl *c, a* | Continental hot spring |
| *Chloroflexus aurantiacus* J-10-fl | Filamentous anoxygenic phototrophs (FAPs) | Bchl *c, a* Bchl *a* | Continental hot spring |
| *Roseiflexus castenholzii* HLO8, Type strain (DSM 13941) | Filamentous anoxygenic phototrophs (FAPs) | Bchl *a* | Continental hot spring |
| *Heliobacterium modesticaldum* Ice1, Type strain (ATCC 51547) | Heliobacteria | Bchl *g* | Continental hot spring |

Abbreviations: *Bchl*, bacteriochlorophyll; *Chl*, chlorophyll.

**Table 2.** Composition of microbial mats analyzed in this study.

| Microbial mat type | Major phototrophic representatives | Pigments | Environmental source |
|---|---|---|---|
| *Mastigocladus* streamers | Cyanobacteria (*Mastigocladus* sp.), anoxygenic phototrophs (*Chloroflexus* sp.) | Carotenoids, Chl *a*; Bchl *c, a* | Continental hot spring |
| *Phormidium* mat | Cyanobacteria (*Phormidium* or *Leptolyngbya* sp.), anoxygenic phototrophs (*Chloroflexus* sp., *Roseiflexus* sp.) | Carotenoids, Chl *a*; Bchl *c, a* | Continental hot spring |
| *O. princeps* mat | Cyanobacteria (*Oscillatoria princeps*) | Carotenoids, phycocyanin, Chl *a*; Bchl *c, a* | Continental hot spring |
| *Synechococcus*-Chloroflexi mat | Cyanobacteria (*Synechococcus sp. Cyanothece sp.*), anoxygenic phototrophs (*Chloroflexus* sp., *Roseiflexus* sp.) | Carotenoids, Chl *a*; Bchl *c, a* | Continental hot spring |
| Chloroflexi red underlayer | Anoxygenic phototrophs (*Chloroflexus* sp., *Roseiflexus* sp.) | Carotenoids, Chl *a*; Bchl *c, a* | Continental hot spring |
| Chocolate Pots *Synechococcus*-Chloroflexi mat | Cyanobacteria (*Synechococcus sp. Cyanothece sp.*), anoxygenic phototrophs (*Chloroflexus* sp., *Roseiflexus* sp.) | Carotenoids, Chl *a*; Bchl *c, a* | Continental hot spring |

Abbreviations: *Bchl*, bacteriochlorophyll; *Chl*, chlorophyll.

**FIGURES AND CAPTIONS**

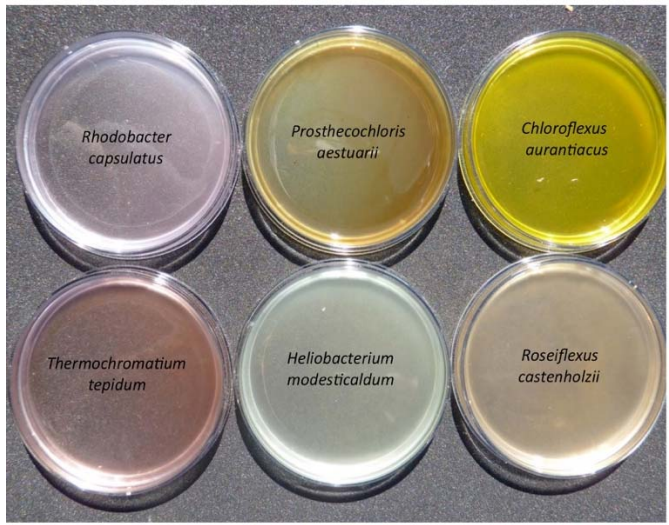

Figure 1.  Pure cultures of anoxygenic phototrophs analyzed in this study.

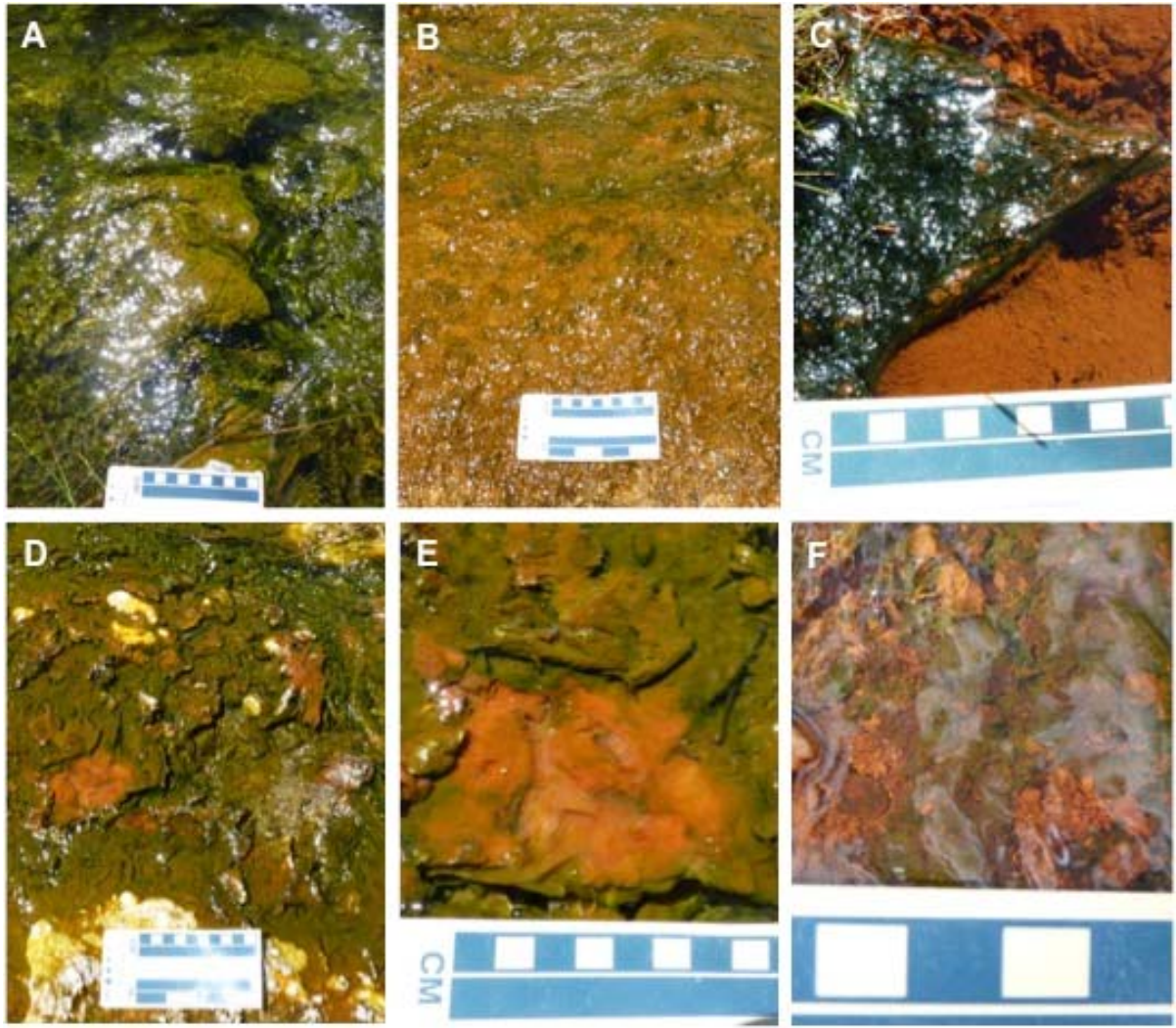

Figure 2. Field photos of photosynthetic mats in the studied locations. A) White Creek *Mastigocladus* streamers, B) White Creek *Phormidium* mat, C) Chocolate Pots *O. princeps* mat on an iron mineral substrate, D) White Creek *Synechococcus*-Chloroflexi mat, E) White Creek Chloroflexi red underlayer, F) Chocolate Pots *Synechococcus*-Chloroflexi mat on an iron mineral substrate. All other mats are on a siliceous sinter substrate.

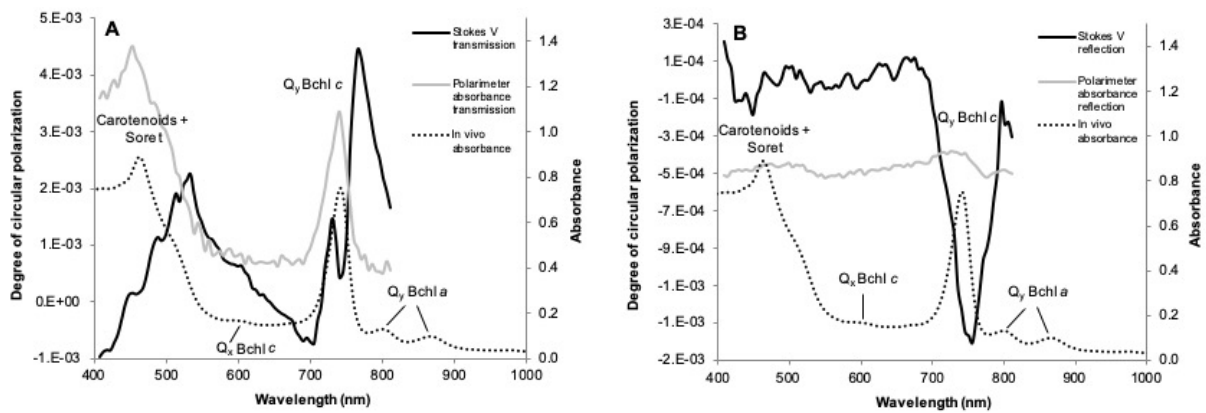

Figure 3. Panel showing correlation between the *in vivo* absorbance, polarimeter absorbance, and degree of circular polarization (Stokes V) for the filamentous anoxygenic phototroph (FAP) *Chloroflexus aurantiacus* (a) in transmission and (b) in reflection. The alignment of the absorption maxima of the $Q_x$ and $Q_y$ bands of bacteriochlorophyll *c* (Bchl *c*) and the circular polarization signals are visible.

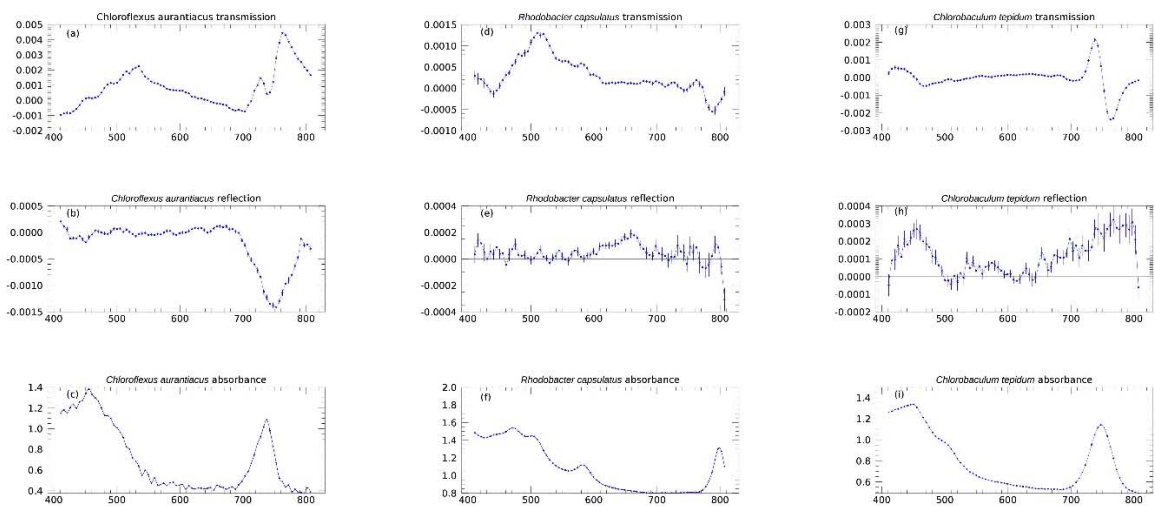

Figure 4. Transmission and reflection circular polarization spectra and absorbance measurements of pure cultures of anoxygenic phototrophs. The Y axis is the degree of circular polarization for the upper two rows, and the lower row shows the absorbance. The X axis is the wavelength (nm). The upper row shows the transmission circular polarization spectra, middle row the reflection circular polarization spectra, and lower row the absorbance measurement.

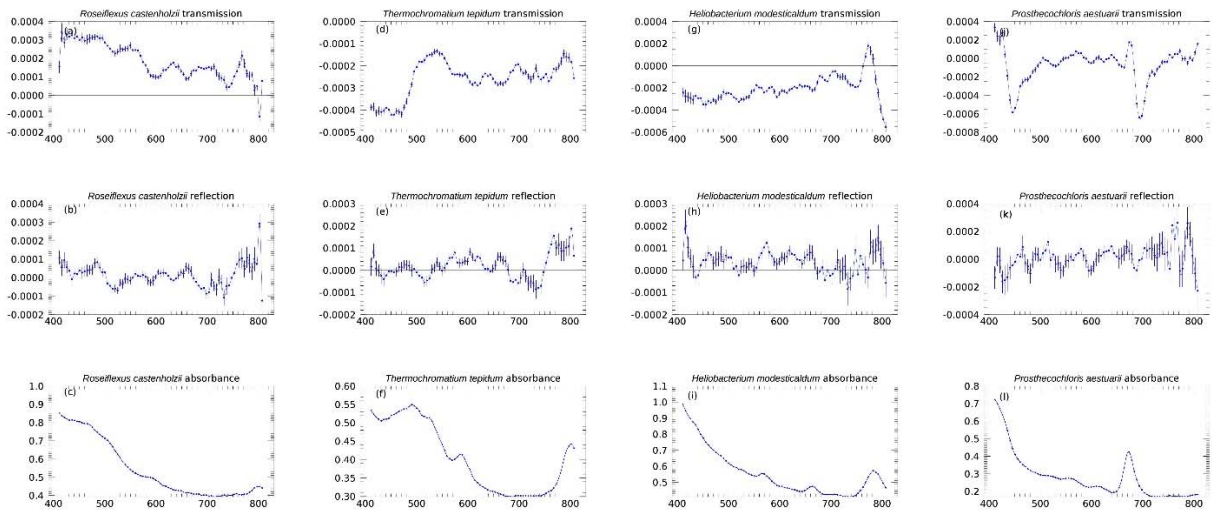

Figure 5. Transmission and reflection spectra and absorbance measurements of pure cultures of anoxygenic phototroph (continued). The Y axis is the degree of circular polarization for the upper two rows, and the lower row shows the absorbance. The X axis is the wavelength (nm). The upper row shows the transmission circular polarization spectra, middle row the reflection circular polarization spectra, and lower row the absorbance measurement.

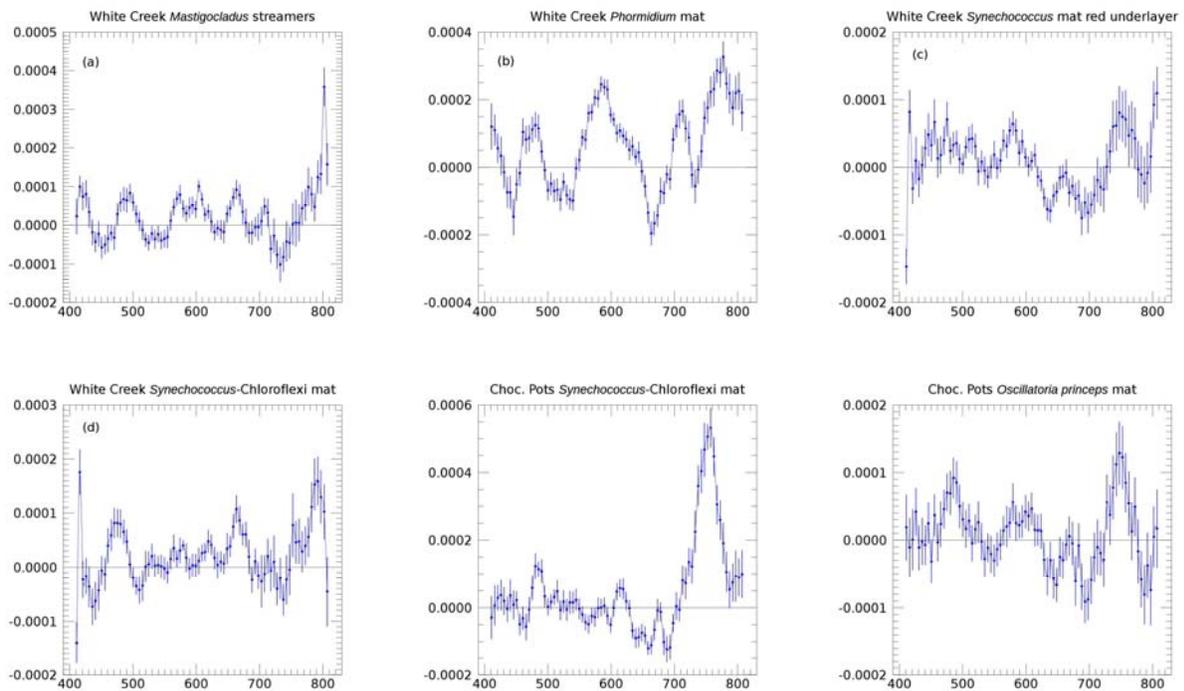

Figure 6. Reflection circular polarization spectra of environmental photosynthetic microbial mats. The Y axis is the degree of circular polarization, while the X axis is the wavelength (nm).

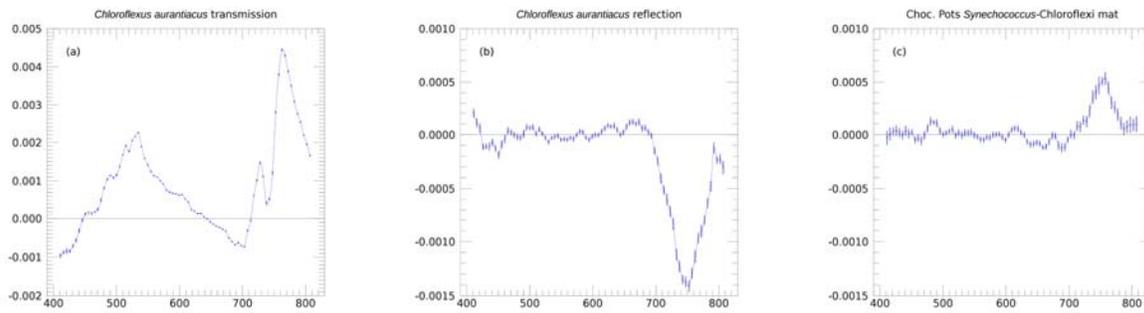

Figure 7. The (a) transmission and (b) reflection circular polarization spectra of the pure culture *Chloroflexus auranticus*, which serve as end member spectra for helping to deconvolve the (c) circular polarization spectra of the complex community in the *Synechococcus*-Chloroflexi photosynthetic mat from Chocolate Pots hot springs. The Y axis is the degree of circular polarization, while the X axis is the wavelength (nm).

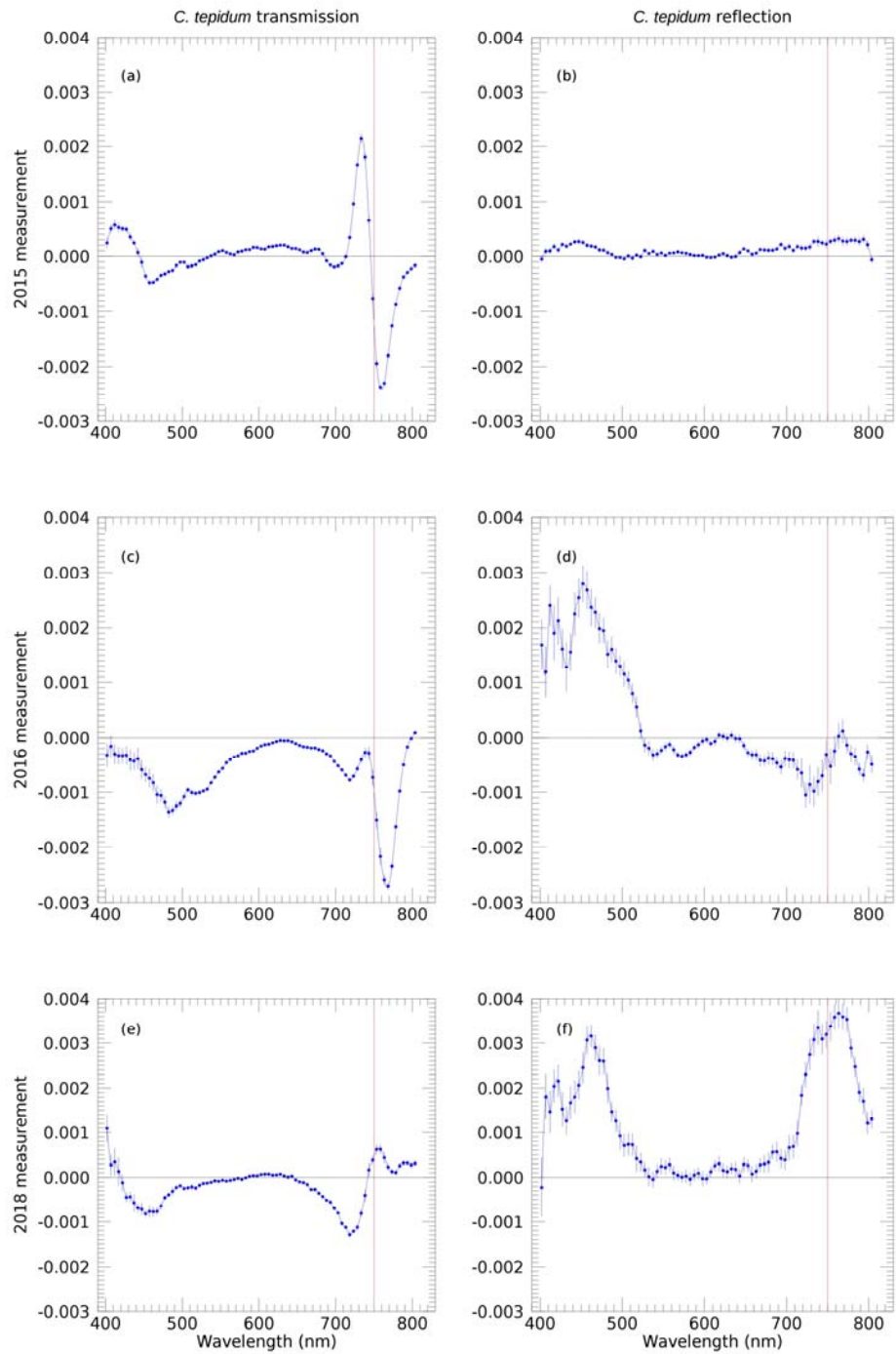

Figure 8. Time course measurements revealing change in the transmission and reflection circular polarization spectra of the green sulfur anoxygenic phototroph *Chlorobaculum tepidum*. Surprisingly, the reflection spectrum of the *C. tepidum* culture was strengthened (f) as the cells lysed and the chlorosomes containing the bacteriochlorophyll *c* pigments were released. The Y

axis is the degree of circular polarization, and the red (vertical) line indicates 750 nm, the wavelength of the Qy peak of Bchl c in vivo, which can occur in the range 740-755 nm.